# Exoplanet Detection with Microlensing

Author: Dr. Aparna Bhattacharya
Department of Astronomy, University of Maryland, College Park
NASA Goddard Space Flight Center

Abstract:

Contents

**Learning Objectives**

- Why is microlensing needed to detect solar system analog planets?
- What is microlensing?
- What ground based surveys use this method today?

- Why high resolution needed for mass measurement of planets?
- What is the future of space survey with microlensing?

**Nomenclature**

| | |
|---|---|
| TESS | Transiting Exoplanet Survey Satellite |
| HST | Hubble Space Telescope |
| LGS AO | Laser Guide Star Adaptive Optics |
| RGES | Roman Galactic Exoplanet Survey |
| GBTDS | Galactic Bulge Time Domain Survey |
| MOA | Microlensing Observation in Astrophysics |
| OGLE | Optical Gravitational Lensing Experiment |
| KMTNET | Korea Microlensing Telescope Network |
| HJD | Heliocentric Julian Day |

## 1. Introduction: Why Microlensing?

The past decade has seen an explosion of exoplanet discoveries. Thousands of exoplanet discoveries have resulted from Kepler and TESS exoplanet transit missions. Doppler radial velocity and ground-based transit surveys have also contributed substantially to the >5500 confirmed exoplanet discoveries. Despite the large number of exoplanets known, our knowledge of the distribution of exoplanets is rather limited. According to the leading core accretion theory, planet formation is most efficient beyond the "snowline", where ices can condense into solid form. The presence of ices increases the density of solids by a factor of a few, and this catalyzes the initial steps of the planet formation process. Planets beyond the snowline are also thought to have important implications for potentially habitable planets in inner orbits as they influence both the formation and the water content of the terrestrial planets. But exoplanet detection beyond the snow line is nearly impossible with the transit method and difficult with the radial velocity method, particularly for M dwarf stars. While the vast majority of the known exoplanets have been discovered by transit and radial velocity methods, these methods can barely probe the primary planet birthplace for the most common stars in the galaxy.

Microlensing is the only method that has been able to probe low mass planets (planets less than a few Jupiter Mass) at large orbital separations. The vast majority of exoplanets found by the microlensing method orbit M-dwarfs beyond the snow line. Most of the planets discovered by transit have orbital separations much smaller than 1 AU because of the high sensitivity of transit searches to the planets at small separations. Figure 1.1 shows that transit photometry can easily detect warm planets close to their host stars. When a planet transits or crosses in front of its host star, the brightness of the host star drops by a small amount, depending on the relative size of the host and the planet. From measuring the drop in the brightness, the mass and radius of the exoplanets can be computed . But for cold rocky planets or gas giants, the orbital period is very high. For Neptune, the orbital period is 165 years. Hence, a Neptune type planet will take more than 100 years to come in front of the host star along our line of sight. We will not be able to measure the orbital period of this planet with a transit unless we decide to wait another 100 years or more. Also, the transit method is efficient for determining exoplanets in nearby systems, which means it is effective in finding planetary systems at a distance of only a few hundred parsecs away from the solar system.

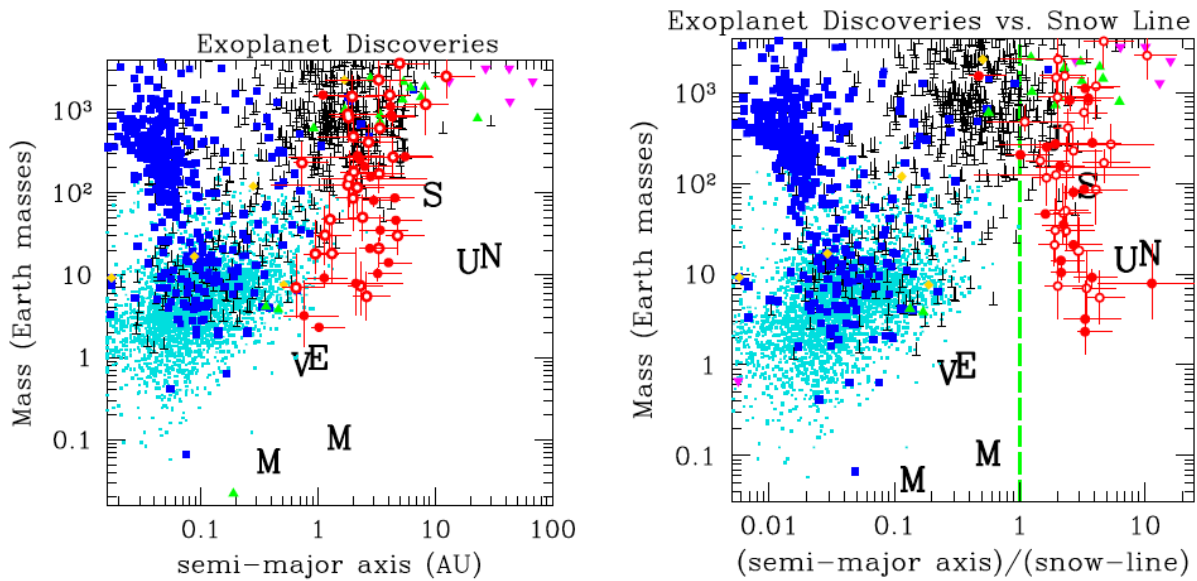

Fig. 1. Distribution of known exoplanet masses plotted vs. semi-major axis (left) and semi-major axis normalized to the location of the snow-line (right) at $a_{snow}$ = 2.7AU M/$M_{Sun}$ (green dashed line). Planets found by the Doppler/RV method (black lines with upward error bars indicating the sin i uncertainty, where i refers to line-of-sight inclination), transits (blue squares), microlensing (red error bar crosses), direct detection, (magenta triangles) timing (green triangles) and transit timing variations (gold diamonds). Small cyan spots indicate Kepler planets and planet candidates without mass measurements. Black letters show the locations of our Solar System's planets. Microlensing discoveries with direct host star and planet mass measurements are indicated with solid red dots.

The other popular method for detecting exoplanets is Doppler radial velocity method. When a planet orbits its parent star, the host star's radial velocity with respect to earth varies due to the gravitational pull of the planet. As a result of this variation, we can see the Doppler shift in the spectrum of the star. This shift can be measured and modeled with binary mass function to obtain the planetary parameter M*sini*. M is the planet mass and i is the inclination of the planet orbit with respect to its host star. With current spectrographs we can measure radial velocities as low as 1 m/sec variation in the spectra. But the variation in radial velocity of the star depends on the relative size of the planet and host star and the separation between them. The higher the planet-star separation and the lower the mass of the planet, the lower is the gravitational pull and hence the lower the variation in the radial velocity. As a result, the Doppler shift in the spectrum is small and very difficult to measure. An alternative method for detecting exoplanets is by microlensing method.

## 1.1 What is Microlensing?

Microlensing is unique in its sensitivity to cold low mass planets at large separations. When a star crosses another star along the line of sight, the background star is gravitationally lensed by the foreground star. The background star is known as the source and the foreground star is known as the lens. Due to this gravitational lensing, the source star brightness is magnified, and we observe a Paczyn´ski (Fig. 2 and 3) light curve. If the lens

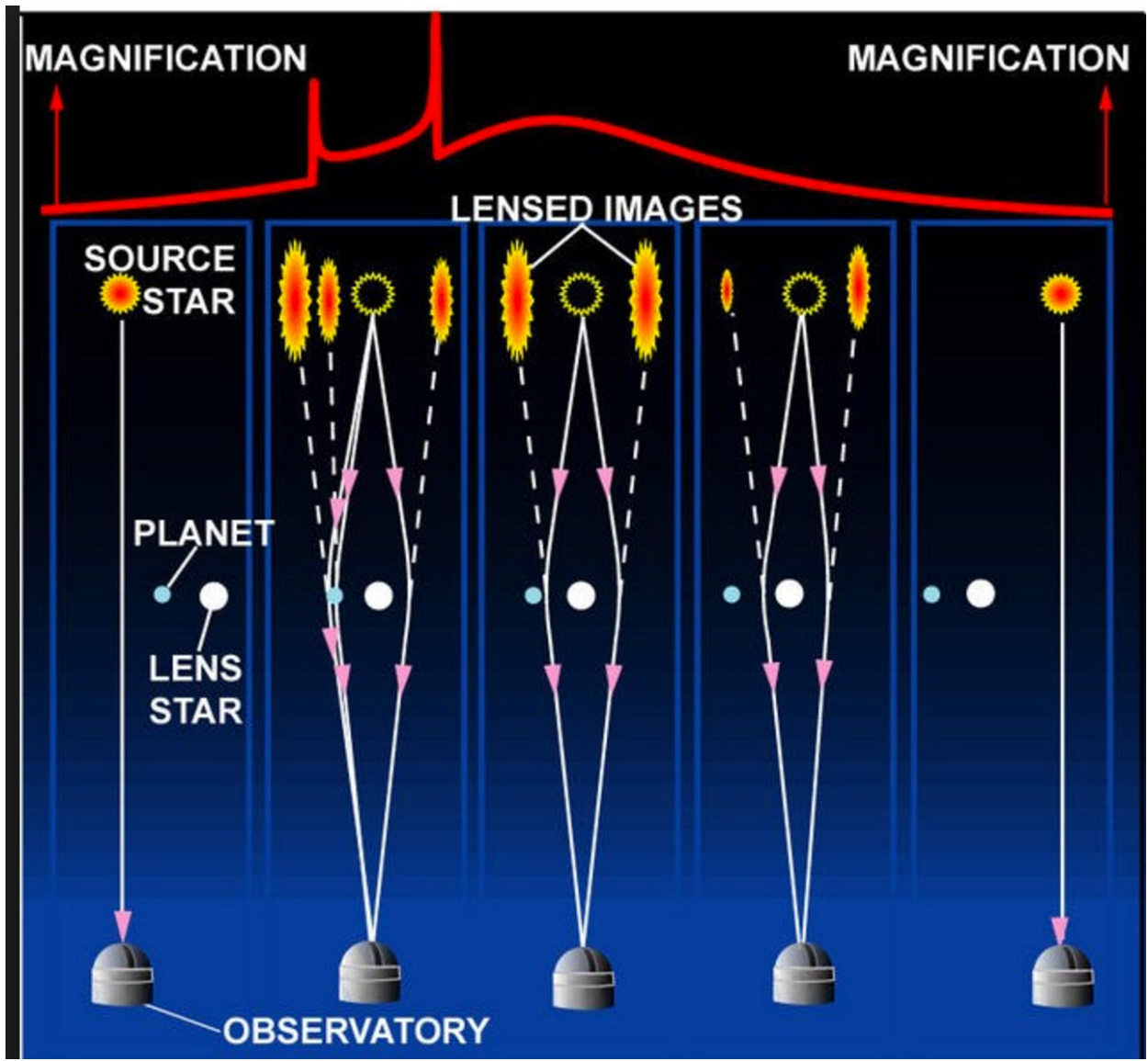

Fig. 2 : The background ``source" star is lensed by the foreground ``lens" system. Here the lens system consists of two lens bodies: the primary lens body is the host star and the secondary lens body is the planet. Due to the lensing effect, multiple images of the source

star are created. But these images are separated from each other on the sky by less than milliarcseconds. Hence, we cannot see these images separately. Instead, we see a combination of all these images. The top panel shows the light curve of the source with its brightness being magnified. When the light from the source is gravitationally lensed by the planet, the planet signature appears in the light curve, as shown in the top panel of the figure.

star happens to host a planet, the source is gravitationally lensed by the planet as well and the planet signature is observed in the light curve. By modeling the light curve with the binary lens model, we can determine the planet-host star mass ratio and estimate the planet-host star masses and their distance from Earth. With the high resolution follow-up images, we can measure the mass and the orbital separation of the planet and the host star and their distance from Earth. This method has discovered about 210 planetary systems. Since the detection of the planet does not depend on the light from its host star, it is possible to detect the planetary systems at a distance as large as 8 kpc, close to the Galactic center.

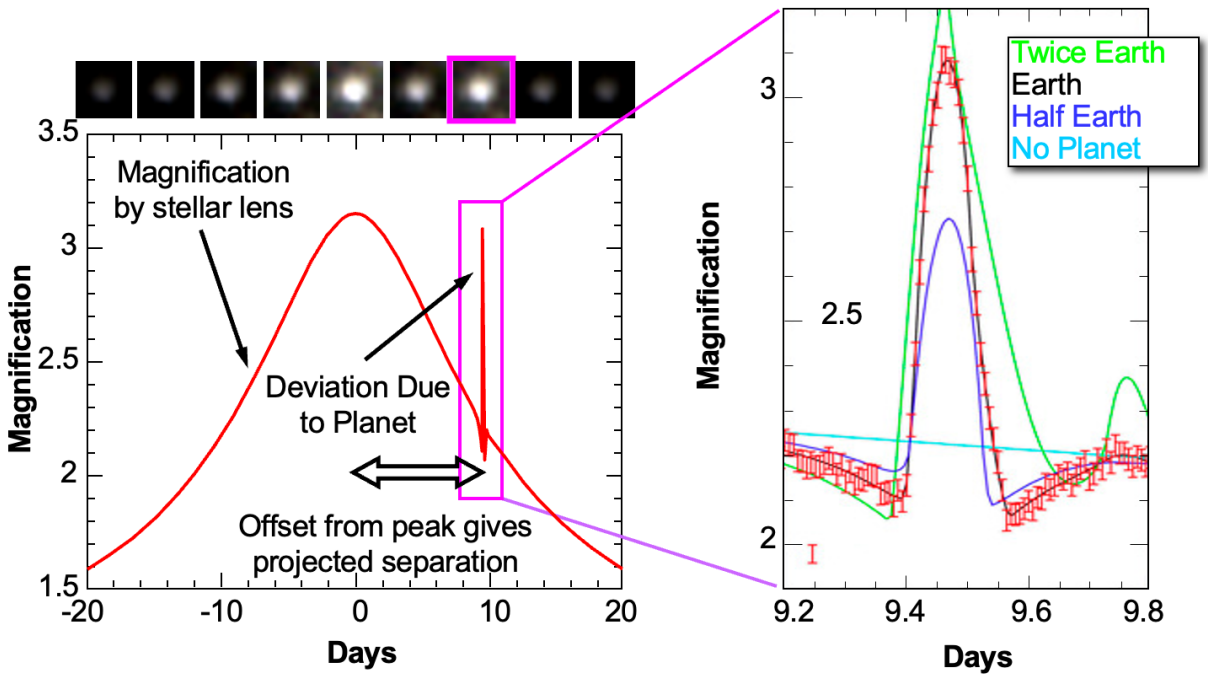

Fig. 3 : The photometry of a microlensing anomaly signature. As shown in top left panel, as the lens comes in front of the source, the source brightness is magnified and over time we notice first an increase in brightness followed by a decrease. This time series feature was first noticed by Paczynski and hence known as Paczynski light curve. Also, if a planet

happens to be at a distance of 2-3 AU from the lens star, the gravitational field of the planet will also cause distortion in the path of light and show a feature which we call planetary anomaly. As shown in the right panel, if the planet mass is higher, the anomaly signal from the planet is also larger.

## 1.2 Where do we see microlensing exoplanets?

The microlensing phenomenon is quite rare. The stellar lensing probability is only a few in a million of stars. And the chance of a planetary signal is about 0.001-1 % depending on the architecture of the system. Hence to increase our chance of observing such phenomenon, we observe in the direction of Galactic bulge where there are hundreds of millions of stars. Hence the probability of one star coming in front of another star towards bulge is much higher. Note that even though the events are rare, the signals of the planets are strong and very much detectable. During the event, the source gets highly magnified; the total brightness is dominated by the magnified brightness of the source. As a result, the lens host star is barely detectable during the event.

## 1.3 What kind of exoplanets are discovered using microlensing and why?

If the exoplanet is located about 2-3 AU from the host star, it has the peak sensitivity of being detected. For this reason, the planets detected using microlensing method are located around the habitable zone and likely are ice giants (Neptune, Uranus) or Jupiter- or Saturn-analog planets.

The planets discovered by microlensing are usually 1-8 kpc distance away from earth. These distances are too great to enable the atmospheric study of these planets with current technology. However, since we do not need to detect the light from the host star to detect a microlensed planet, we can detect rare planets around white dwarfs, neutron stars and black holes or planets with no host stars.

## 1.4 How long do microlensing events last?

A typical microlensing stellar event lasts 30-40 days and a planetary anomaly lasts several hours to few days. However, if the event is a free-floating planet, then we will only observe a Paczynski curve for a short duration of 1-6 hours. If the lens is a black hole, we might see the Paczynski curve for a few months to few years.

## 2 Free-Floating Planets

Microlensing is currently the only method that is capable of detecting free-floating planets over a large range of distances from Earth. Since the detection of the planet does not depend on the light from its host star, it is possible to detect isolated (no host) planetary systems at a distance as large as 8 kpc, close to the Galactic center. These planetary systems are ``isolated" because their light curves do not reveal any evidence of the host star. It is highly possible that these are rogue planets [Sumi et al. 2023]. The same paper predicts a higher abundance of Jupiter mass objects. Their analyses indicate that, for each main-sequence star, there are about ≈20 Jupiter-mass objects. It seems likely that a high fraction of these could really be free-floating planets that have become unbound from their host stars. Such planets are expected to form from a variety of mechanisms, but it seems difficult for the bulk of this new population to be explained by any of these mechanisms alone. One possible explanation is perhaps they were ejected from binary star systems.

This discovery of unexpectedly large number of free-floating planets has led to the suggestion that a large fraction of this newly discovered planetary mass population may be bound to stars. Whether these are wide-orbit bound planets or free-floating planets that still needs to be verified with high resolution images. If high resolution images taken with Keck or Hubble Space Telescope shows a host star with a planet bound at a very far distance from the host, then that would confirm that they are not free-floating planets. But currently microlensing is the only method that is capable of discovering free-floating planet candidates.

## 3    Current Ground Based Microlensing Surveys

There are mainly three microlensing survey telescope groups: *OGLE*, *MOA* and *KMTNET*. *KMTNET* is a recent addition to this group. These telescopes around the globe monitor several bulge fields each night. mainly during the "bulge season" between February and November. Every night the microlensing alert systems of these groups issue an alert when a potential detection occurs due to an anomaly from the single lens Paczynski light curve. Whoever notices the anomaly immediately alerts all of the microlensing follow-up groups, who quickly attempt to monitor the event in real time

There are four main microlensing follow up groups: PLANET, ROBONET, MINDSTEP and μFUN (MICROFUN). These groups perform follow up observations with telescopes from different parts of the world, mainly located in Cape Town, South Africa; Perth, Australia; Las Campanas, Chile; Auckland, New Zealand and Hawaii, USA. Note, most of these telescopes are in southern hemisphere because from the southern hemisphere we can observe the Galactic bulge for a longer time every night. The bulge can be observed from Hawaii for ~4 months. The anomalies in the light curves are followed carefully by these groups. Later, the data of both survey and follow-up telescopes are reduced. These reduced data are used for the light curve fitting.

The *OGLE* (Optical Gravitational Lensing Experiment) collaboration operates a microlensing survey towards the Galactic bulge, with the 1.3 meter Warsaw telescope from Las Campanas observatory in Chile. Most of the OGLE-IV (phase 4) observations were taken in the Cousins I-band, with occasional observations in Johnson V-band. Currently the *OGLE* Early Warning System (EWS) alerts more than 2000 microlensing candidates each year. The *MOA* (Microlensing Observation in Astrophysics) collaboration also operates a microlensing survey towards the Galactic bulge with the 1.8-meter MOA-II telescope from Mount John Observatory at Lake Tekapo, New Zealand. The observations are mostly taken in the MOA-red wide band field, which covers the wavelengths of the standard Cousins R + I bands. Currently MOA reports about over 500 microlensing events each year.

Both of these telescopes have relatively large fields-of-view, 2.2 deg² for MOA-II, and 1.4 deg² for OGLE-IV. These enable survey observations with cadences as high as one observation every 15 minutes, and this allows the surveys to detect the sharp light curve features of planetary light curves, when they are only smoothed by the finite source effects of a main-sequence source star. It is the high cadence observation of microlensing events of MOA-II and OGLE-IV survey that helps in detecting microlensing anomalies, including the microlensing planetary signatures.

## 4   Measurement of the Mass of the Host and Planet and their Distance

The modeling of the ground-based light curves only yields the planet-host mass ratio, but not the physical masses or distances to the system. To obtain the masses, we need to detect the lens system and measure its brightness. Remember that the lens host is barely visible in ground-based observations, especially during the event. Even after the event, it is not possible to detect the lens host star separately from the source. There are three equations that are important for measuring the mass of the planets and the host.

$$M_L = \frac{c^2}{4G} \theta_E^2 \frac{D_S D_L}{D_S - D_L} \quad (1)$$

$$M_L = \frac{c^2}{4G} \tilde{r}_E^2 \frac{D_S - D_L}{D_S D_L} \quad (2)$$

$$M_L = \frac{c^2}{4G} \tilde{r}_E \theta_E \quad (3)$$

In the above equations, the mass of lens is $M_L$, the distance to the source is $D_S$, the distance to the lens is $D_L$, and $\theta_E$ is the angular separation of the images of the source and the lens on the sky when the point lens cross in front of the point source – this parameter is often referred to as angular Einstein ring radius. $\tilde{r}_E$ is the projected Einstein radius. Here is a cartoon to show the geometric structure and the meaning of the above mentioned parameters, based on Gaudi 2010:

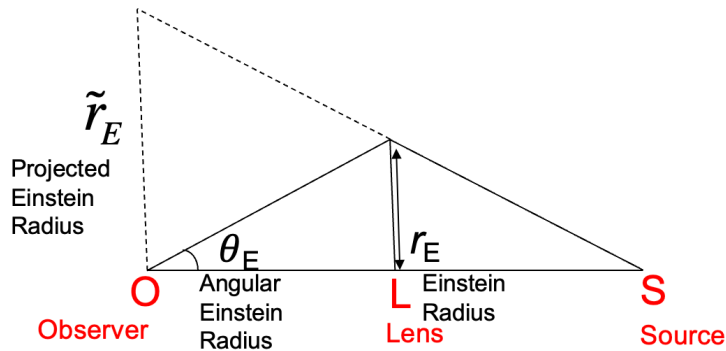

The microlensing parallax is denoted by:

$$\pi_E = \frac{1}{\tilde{r}_E} \qquad (4)$$

We usually know the distance to the source from Galactic bulge-based Bayesian models, and because we know that the source is at the center of the Galactic bulge at a distance of about 8 kpc. If we can measure angular Einstein radius and the microlensing parallax from the light curves or time series data like in Figure 3, then we can solve equations 1 and 2 and obtain the mass of the lens and the distance to the lens. Since we already know the planet-host mass ratio from the light curve analysis, we can then obtain the planet mass by multiplying $M_L$ with $q$.

### 4.1 Microlensing Parallax

There are three different ways we can measure the microlensing parallax.

### 4.1.1 Orbital Parallax

Orbital microlensing parallax arises due to the earth's orbital motion. When the earth is rotating around the sun, the acceleration due to this rotation adds a modification to the light curve of the source brightness. This signature is negligible if the event is on thei order of days or less. But if the event is on the order of a month or more then this signature is prominent. Also, during autumn and spring, earth's acceleration is higher and this signature is comparatively more prominent than during the other seasons. For parallax effects, there are two more nonlinear parameters are added to the list of fitting parameters: $\pi_E = \pi_{rel}/\theta_E$ ( the lens-source relative parallax in units of Einstein angular radius) and a direction of the relative lens-source proper motion relative to north towards east, $\varphi_E$.

Here is an example from Muraki et al (2011) that demonstrates microlensing orbital parallax:

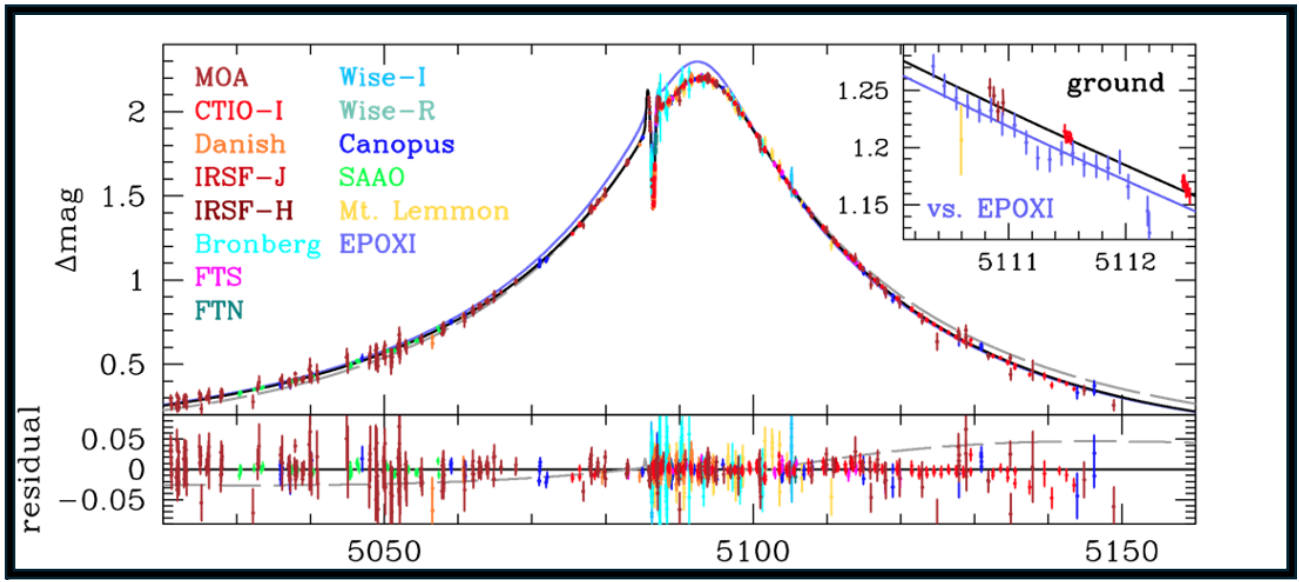

Figure 4: Data and best-fit model of the MOA-2009-BLG-266 microlensing event plotted with respect to magnitude of the unmagnified source. The x axis shows the time in unit of HJD-2450000 days. HJD is Heliocentric Julian Day. As the reader can see, the event occurred over 100 days. During these 100 days earth was also moving in its orbit. This orbital motion of earth caused a microlensing parallax. The upper panel shows the part of the microlensing light curve magnified by more than 25%. The sub-panel at the bottom shows the residual to the best fit model, which is given by the solid black curves. The grey-dashed curve in each panel is the best fit non-parallax microlensing model, which is not the best fit model. The best fit model is the model that detects microlensing orbital parallax – this parallax happens due to the motion of Earth.

### 4.1.2 Terrestrial Parallax

When we observe the same microlensing event from three different locations on Earth, especially separated by time zones, we can see the light curve is moved in the observations of each location. A demonstration with real data is shown below in Figure 5, from Gould et al. 2009.

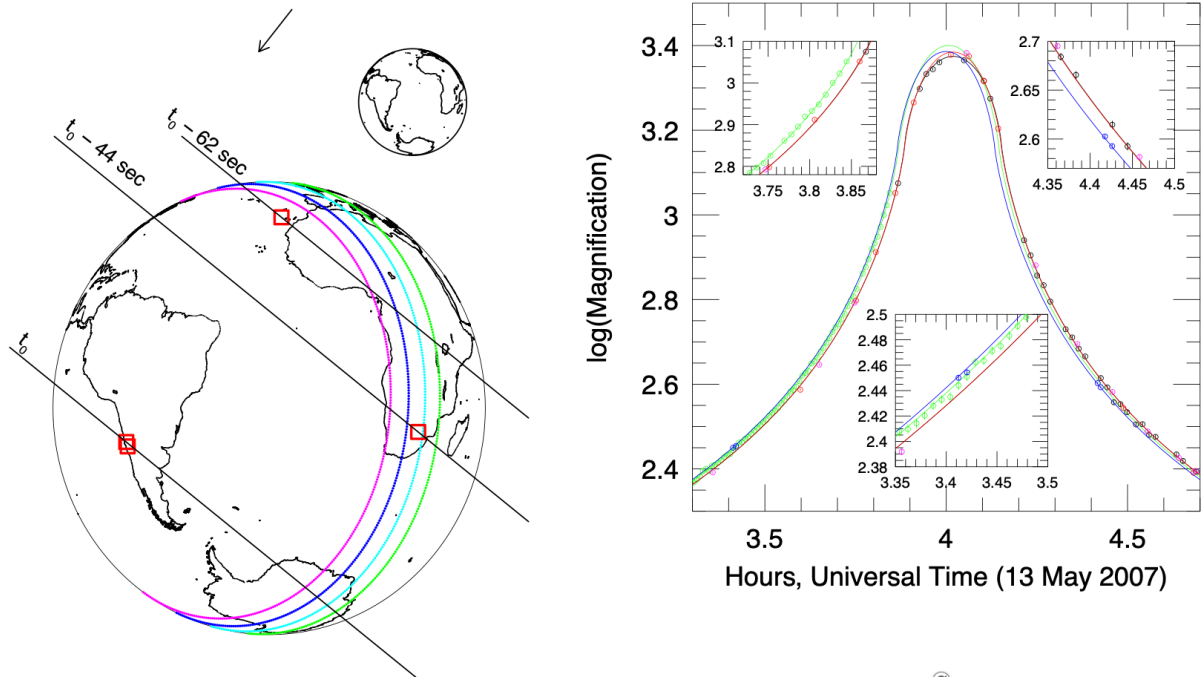

Figure 5: Gould et al. 2009 shows that the observations taken from different sites on Earth can show microlensing parallax or movement of the light curve. Each color – green, cyan, blue and magenta corresponds to a different observatory on Earth. The data from each of those observatories is plotted with the same color as the observatory. As shown, the data from each observatory is shifted. This is called terrestrial parallax- a microlensing parallax due to the motion of Earth.

### 4.1.3 Satellite Parallax

There is another way to measure mass of the host and the planet – satellite parallax. This microlensing parallax occurs when the microlensing event is observed from the Earth and a satellite (as in a spacecraft rather than a moon), like the Earth and *Spitzer*, or from two different satellites, like *Euclid* and *Roman*. Observing the same event from the two different lines of sight means that we can see the lens approaching the source from two different angles. Hence, the peak time of the microlensing event is shifted in the light curves. If we can measure this parallax, then we can also measure the mass of the lens and the distance to the system, which will yield mass of the planet as well.

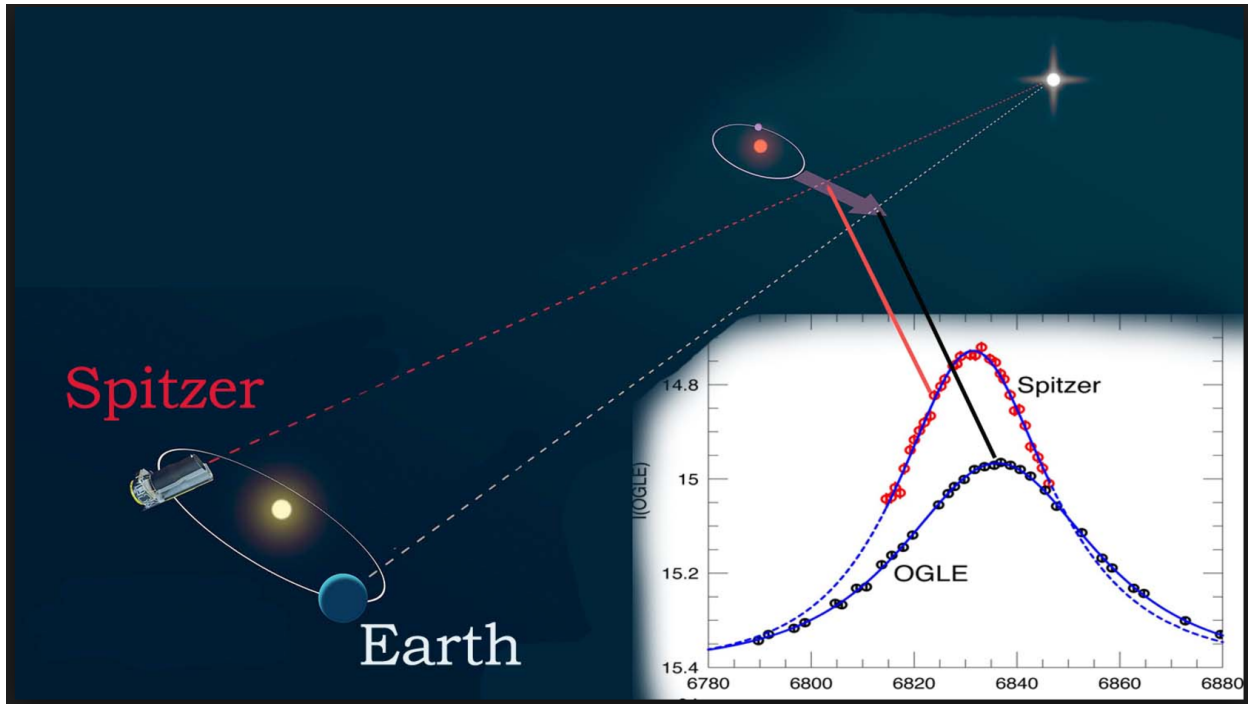

Fig. 6: Satellite parallax from the space-based *Spitzer* telescope and the ground-based *OGLE* telescope of a single lens microlensing event.

If we can detect the microlensing parallax and angular Einstein radius from the light curves, then we can use Equations 1 and 2 to solve for the mass of the planet and the host star. However, from most ground-based survey data, we cannot detect both of these quantities. Most of the time we can only detect one of them. Ground-based surveys use telescopes that are not enabled with sufficient resolution and photometric sensitivity to detect both of these parameters. Hence, to measure mass of the host and the planet, we need to solve a third equation. We can detect the lens in high-resolution observation and use a mass-luminosity relation to get a third mass-distance relation.

### 4.2 Mass Measurement with High Angular Resolution Images

Ground-based survey light curves usually yield one of the two parameters: angular Einstein radius or the microlensing parallax. However, we can still measure the mass of the host and the lens by utilizing a third mass-distance equation: the empirical mass luminosity equation (shown below). If we can detect the lens in high resolution images and measure the brightness, then we can use the following empirical mass luminosity relation to get a third mass-distance equation. Combining this equation with either equation 1 or 2, we can solve for the mass of the lens and host.

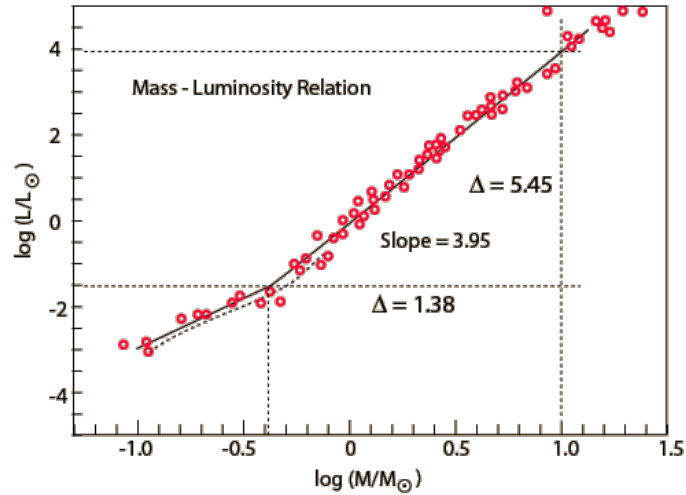

Fig. 7 shows a plot of Mass-Luminosity relation plotted in log scale. This is an example that when we have a graph like this, and we have brightness/ luminosity of the host star from the high resolution image, we can find the corresponding mass using a graph like this.

Microlensing is a rare phenomenon that happens when two stars – a lens and a background source are exactly aligned along the line of sight. During the microlensing, the source is so bright that we cannot see the lens star. But as microlensing is over in a few days or months, the source and lens are separate. During this phase we can refer to the source and lens as "partially resolved". At this time, we can observe the event with high-resolution telescopes. We need high resolution, because normal ground-based telescopes cannot distinguish two stars separated by ~10-60 mas separation. Because the separation is so small we need high angular resolution telescope to see the lens. The following figure shows a comparison of ground based data vs space based high resolution data.

Fig. 8 shows a comparison of crowding of ground-based data vs space-based data. The lens system is typically moving at 5-6 mas/year with respect to the farther source. Hence in 10 years it has moved about 50 mas, which is much below the crowding level of *OGLE*. Hence, we need Hubble Space Telescope or Keck Laser Guide Star AO system to detect this separation (Bennett & Rhie 2002).

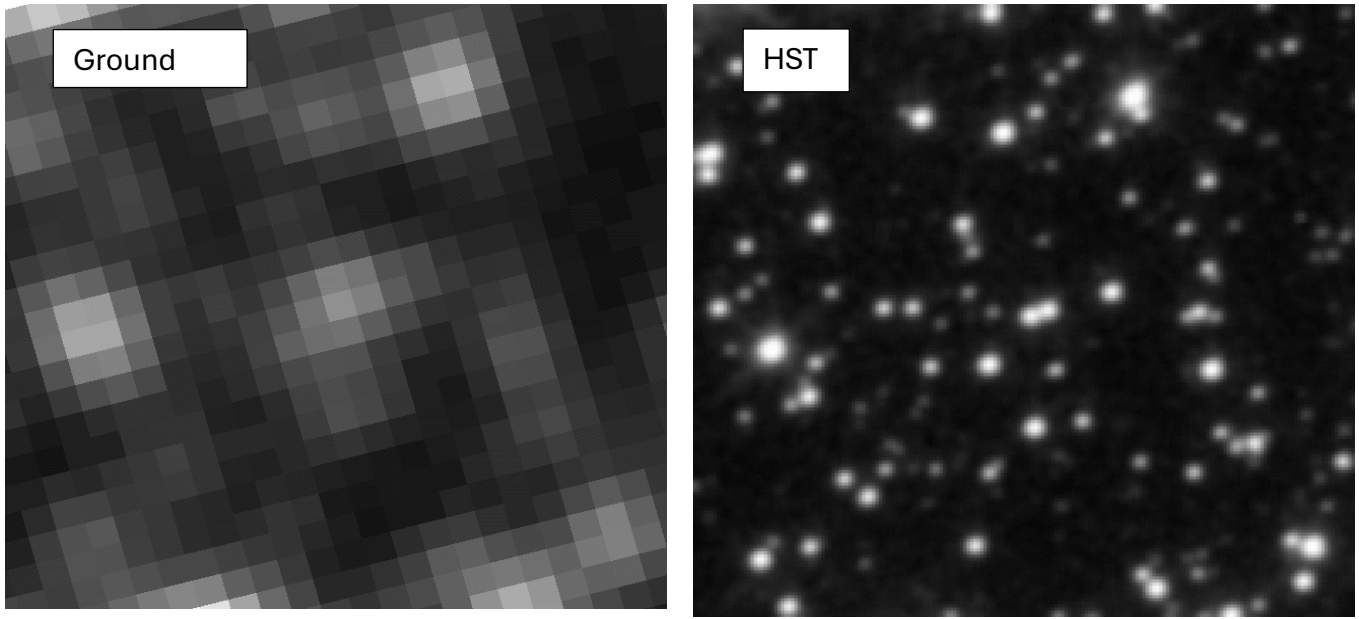

### 4.2.1 Image Elongation

We can observe the lens and source in high-resolution images a few years after the microlensing event. When we observe with a high resolution telescope a few years after the event, we find that the target star is elongated due to the lens and source separating out but not fully resolved- at this stage we will call it "partially resolved". After waiting for long enough , we can actually observe that the lens and source are completely resolved. In this case, if there is no other star nearby, we can identify the lens. Fig. 5 demonstrates the image elongation technique.

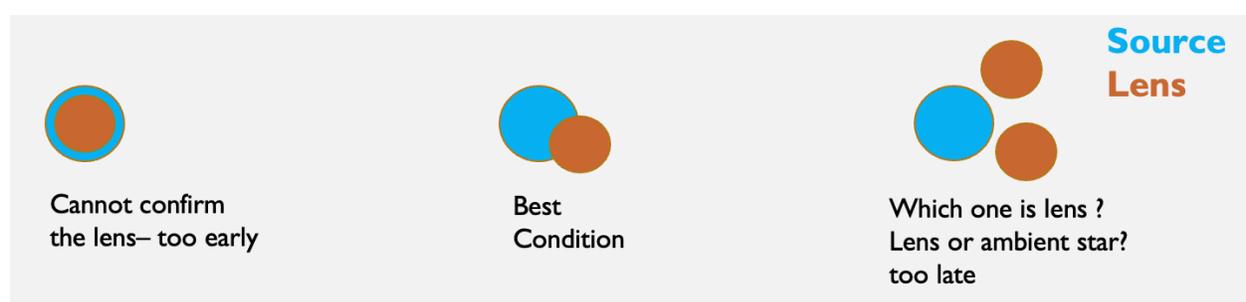

Fig. 9: From left to right shows the timeline: when the lens and source are aligned in the line of sight and the lens is too faint compared to the magnified source. The best condition to detect the lens is when the lens and source are partially resolved – about 5-10 years after the peak of the event, with current resources. Once they have already separated, it can be difficult to detect the lens since there are other unrelated nearby stars.

An example of such a mass measurement obtained with high resolution observations is given in Fig. 10. OGLE-2012-BLG-0950 was an event that was observed with both HST in the V and I bands, and with the Keck LGS AO telescope in the K passband. V is bluer passband and K is the reddest passband. The source and lens in these images are partially resolved. The two stars, lens and source were detected in all 3 passbands independently at same position. The separations measured are consistent. The separation measured yielded both the components of the relative proper motion between lens and the source – this constrained several light curve models. The brightness of the lens also indicated the mass of the lens using the empirical mass-luminosity relations. Together this information helped us to obtain the mass of the host star and its distance from Earth. Then, since we already knew the mass ratio and separation ratio from light curve modeling, we were able to derive the planet mass and projected separations.

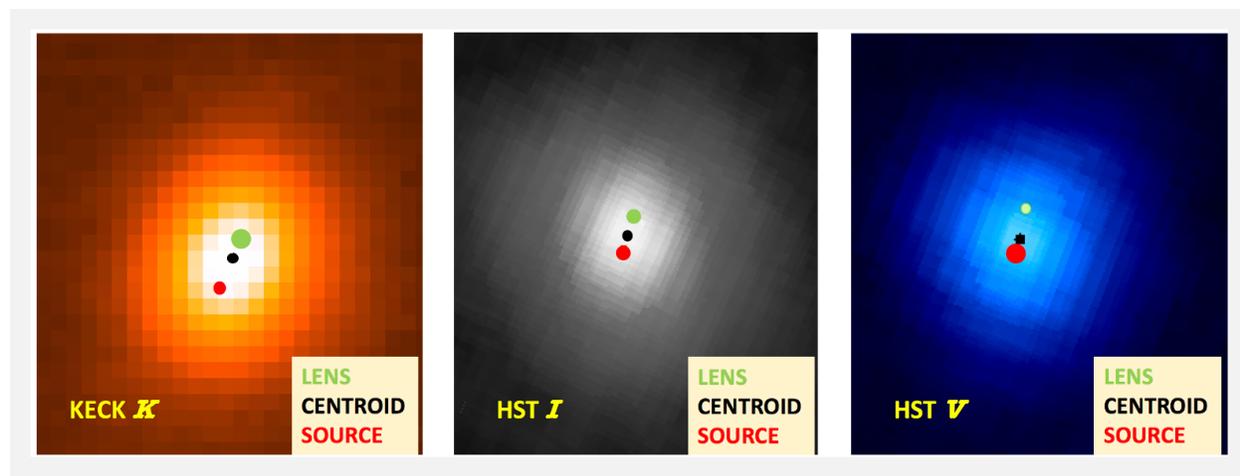

Fig.10: Bhattacharya et al. 2018 shows event OGLE-2012-BLG-0950 observed with red Keck K band with LGS AO and HST I and bluer V bands. The observation was taken 6 years after the peak of the event. The lens and source were separated by 34 mas at that time. The lens and source were detected and measured independently in 3 different passbands. They all yielded consistent results. Since the lens is closer than the source and fainter/redder, the lens is brighter in K band and fainter in V band. This change in flux shows that the centroid of the two stars keep moving from being closer to the lens in red passband to being closer to the source in the bluer passband. Movement of this centroid is known as color-dependent centroid shift.

## 5  Future Space Based Microlensing Surveys

Microlensing is unique in its sensitivity to cold and low-mass planets beyond the snow line. This strength of the microlensing method was recognized in the 2010 New Worlds, New Horizons report, in its selection of an exoplanet microlensing survey for the top-ranked Nancy Grace Roman Space Telescope. Currently it is at 4 billion $USD budget, the second most expensive telescope ever after the James Webb Space Telescope. The field of view of this telescope is 100 times greater than that of the Hubble Space telescope, and yet the resolution of images is comparable to HST. The Roman telescope will launch probably in the Fall 2026, and no later than Summer 2027. This telescope was formerly known as the WFIRST (Wide Field Infra-Red Survey Telescope) mission. According to the WFIRST SDT Interim report and Penny et al. 2019, Roman (formerly WFIRST) will be able to detect 1500+ planets with masses as low as Mars mass and distributed across the galactic bulge. This mass regime is significant because with current technology (as of 2025), detecting sub-Earth mass planets remain challenging and rare.

The Roman microlensing Exoplanet campaign is now known as Roman Galactic Bulge Time Domain Survey (aka RGBTDS). This survey will observe 8 fields in the galactic bulge in the wide Infrared passband F146 every 15 minutes and with two additional filters every 12 hours. There will be 6 seasons of observations of 72 days each, spread-out across 5 years. With the microlensing exoplanet campaign, it will be possible to discover thousands of exoplanets in less explored territory: cold low mass planets beyond the snowline, exo-moons [Bennett et al. 2013] and free-floating planets [Sumi et al. 2023]. The advantage of microlensing is that we don't need to ``see'' the host star to detect the planet. This enables us to detect both bound and unbound planetary systems located near the Galactic bulge. What Kepler has been to warm gas planets, Roman will be to cold rocky planets. Determining the demographics of a large sample of exoplanets will help us to understand and verify different planet formation and evolution theories.

The major advantage of Roman over ground-based microlensing is that due to its high-resolution capability, Roman will be able to directly observe the host stars, leading to the mass measurement of the exoplanets. Measuring the planet mass with the lens detection from the high-resolution images is going to be the primary method for Roman exoplanet mass measurement. This method will help us to build the statistics for the exoplanets as a function of their host star masses and over the distance to the Galactic center. This will fulfill the bigger aim of Roman microlensing mission of completing the exoplanet census. This will answer following science questions:

(1) How common is our solar system?
(2) Did the water on Earth come from the ice giants that are located beyond the snowline?
(3) Did all the planets form beyond snowline and migrated later?
(4) How often do planets get ejected from their host system?

It is possible to obtain also the masses of the free-floating planets by observing them simultaneously with the Euclid and Roman space telescopes. Euclid launched in 2023 to L2 orbit. Hence, the separation between Euclid and Roman will be just right to do a parallax measurement similar to what Spitzer did and obtain masses. However, this observation is still unplanned and under consideration. This could be a crucial method to obtain masses of free-floating planets.

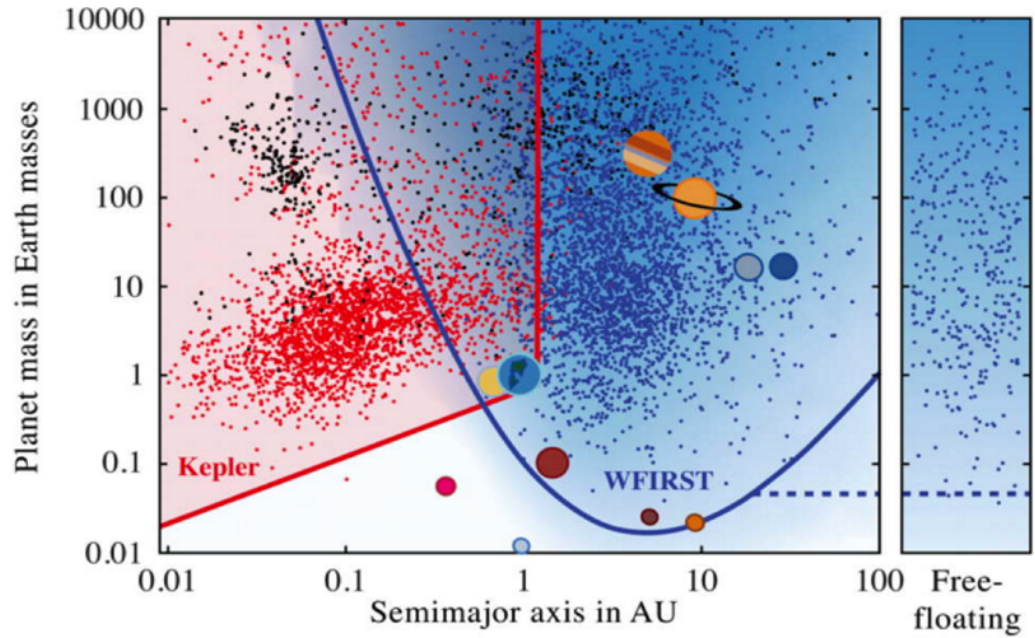

Fig. 11: The colored shaded regions show approximate regions of sensitivity for Kepler (red) and Roman (blue; formerly named WFIRST). The solar system planets are also shown, as well as the Moon, Ganymede, and Titan. Kepler is sensitive to the abundant, hot and warm terrestrial planets with separations less than about 1.5 AU (Penny et al. 2019). On the other hand, Roman is sensitive to Earth-mass planets with separations greater than 1 AU, as well as planets with masses down to roughly twice the mass of the moon at slightly larger separations. Roman is also sensitive to unbound planets with masses as low as Mars. The small black points show planets detected by radial velocity, ground-based transits, direct imaging, and microlensing surveys (i.e., all known exoplanets not found by Kepler). The small red points show candidate planets from Kepler, whereas the small blue points show simulated detections by Roman; the number of such discoveries will be large, with roughly 1600 bound and hundreds of free-floating
planet discoveries. Thus, Roman and Kepler complement each other, and together they cover the entire planet discovery space in mass and orbital separation, providing the comprehensive understanding of exoplanet demographics necessary to fully understand the formation and evolution of planetary systems. Furthermore, the large area of Roman discovery space combined with the large number of detections essentially guarantees a number of unexpected and surprising discoveries.

## 6  Future ground based telescopes taking microlensing observations

PRIME – **Pr**ime focused **I**nfrared **M**icrolensing **E**xperiment Telescope is a 1.8 m telescope located in Sutherland, South Africa. It is going to monitor Galactic bulge fields every year in survey mode starting from 2024 officially. This telescope is using the same H4RG infrared detector as Roman Space Telescope. This telescope is mostly funded by Japan Government.

Vera C. Rubin Observatory – which was previously known as **Large Synoptic Survey Telescope** (**LSST**), is an astronomical observatory under construction in Chile. This observatory will observe in 320–1060 nm. It was the top recommendation for large ground-based observatory according to 2010 decadal survey. Simultaneous observation with Rubin and Roman will help to measure the parallax (satellite parallax in this case) of several microlensing events including the free-floating planet events. This telescope is funded by DOE – Department of Energy USA.

**Conclusions**

Microlensing is the only method through which one can detect wide-orbit low-mass exoplanets. Even though it has only detected 210 planets in last 25 years, the coming 10 years is going to be revolutionary for microlensing exoplanet studies. The launch of Nancy Grace Roman Telescope will yield the discovery of thousands of solar system-like planets and begin a new era in exoplanet studies. Everything we have learned so far and all the tools described in this chapter is a preparation towards this study.

Nevertheless, already some interesting planet formation information has arisen out of the existing microlensing data. In Suzuki et al 2016 and Suzuki et al 2018 studies have shown that there is an abundance of sub-Saturn planets in our galaxy. However, planet formation theory predicts a lack of this kind of planet. This discrepancy is still unresolved. Microlensing has also detected the first Jovian analog planet with a mass of about 5 Jupiter masses orbiting a white dwarf at about half the distance of Jupiter orbit (Blackman et al. 2021). This is a remarkable discovery because when stars go through a phase of evolution into Red Giant Phase and then become White dwarf, most exoplanets around them are thought to be destroyed. However, this Jovian analog planet (Blackman et al. 2021) seems to exist around the white dwarf host, which brings up questions like, 1. did this planet survive the stellar evolution phase? or 2. did it migrate from a different orbit or 3. was it formed around the White Dwarf? In the era of Roman there will be a huge set of data that will reveal many more interesting problems like this. We are entering a golden age for studies of solar system-like exoplanets.